\documentclass{article}
\usepackage{epsfig}

\begin{document}
\title{Charged Black Holes with Yang--Mills Hair and Their Thermodynamics
}
\author{Takuya Maki\\
Japan Women's College of Physical Education\\
Setagaya,~Tokyo 157-8565,~Japan\\
Kiyoshi Shiraishi\\
Faculty of Science, Yamaguchi University\\
Yamaguchi-shi,
Yamaguchi 753-8512,~Japan\\
and\\
Satoru Hirenzaki\\
Department of Physics, Nara Women's University\\
Nara 630-8506,~Japan
}

\date{\today}

\maketitle
\begin{abstract}
We present a new class of the black hole solutions of
Einstein--Maxwell-Yang--Mills theory.
These solutions have
both $U(1)$ charge and Yang--Mills hair. We also investigate the
thermodynamical properties.  
\end{abstract}
\section{Introduction}

Black hole solutions play important roles not only in cosmology and astrophysics,
but also in a clear understanding to quantum gravity.
Black holes and the quantum physics have been studied by many authors 
and developed to paradigms `no-hair conjecture' in 
black hole thermodynamics and early stage of the Universe.
These early investigations have been made for simple theories such as
Einstein--Maxwell theory.
The black hole solution with non-trivial configuration of Yang-Mills
gauge fields were 
found by Volkov and Gal'tsov \cite{VG} and Bizon \cite{B} in
Einstein--Yang--Mills (EYM) theory (called `colored black holes' here).

At first sight, this discovery is surprising because there is no analogous
one in Einstein--Maxwell theory.
Their stability and thermodynamics were discussed in connection with the
no-hair conjecture.
It has been pointed out that the solutions are unstable~\cite{SZ,GV} for
the radial linear perturbation and they were interpreted
as sphalerons of EYM theory~\cite{G,MW}. 
After the discovery of the particle-like spherical solution in EYM
theory~\cite{BM},  black hole solutions with non-Abelian hair have
eagerly been researched. Also, the structure of the black holes has
widely been examined.
In similar systems, 
Skyrme black holes~\cite{LM,DHS}, monopole black holes~\cite{LNW}, 
black holes in the theory coupled to Higgs field~\cite{GMO} 
or a dilaton field~\cite{LD} etc. have
been investigated. Maeda {\it et al.} suggested that these black holes
have some universal properties due to the non-Abelian fields and the
stabilities was discussed from a catastrophe theoretical analysis 
of the black hole entropy~\cite{M}. 

In this paper we investigate the black hole solutions of the EYM theory.
The gauge fields coupled to gravity may arise more naturally from
fundamental physics, for example, string theory.
We present and discussed this charged black hole with Yang-Mills hair.
We are interested in the thermodynamics from aspects of the quantum
physics. It is expected that the results give some implications to black
hole thermodynamics. 


In the next section, colored black holes found by other authors are
briefly reviewed to be compared with ones found by us. The thermodynamic
properties are discussed in Sec.~3. Then we give the inverse temperature
versus the entropy-mass diagram. The final section is devoted to the
conclusion and discussions.  

\section{Charged Black Hole with Yang-Mills Hair}
Before proceeding to the black hole solutions in the theory, 
we summarize colored black hole, namely, a discrete family of 
spherically symmetric solutions
numerically found by Volkov and Gal'tsov \cite{VG} and Bizon~\cite{B} in
Einstein--$SU(2)$ Yang--Mills theory. This is the simplest example of
black holes with non-Abelian hair. The black hole solutions can be
obtained by imposing  the spherically symmetric static ansatz for the
metric as
\begin{equation}
ds^2=-fe^{-2\delta(r)}dt^2+f^{-1}dr^2+r^2 d\Omega ^{2}\,,
\label{metric1}
\end{equation}
where 
\begin{equation}
f=1-\frac{2m(r)}{r}\,,
\label{gauge0}
\end{equation}
and 't~Hooft ansatz for the Yang-Mills connection as
\begin{equation}
dA=\frac{1-w(r)}{2g} U dU^{-1}\,,
\label{gauge1}
\end{equation}
where $g$ is the Yang-Mills coupling constant and
$U=\exp(i\frac{\pi }{2}{\tau}\cdot {\bf n} )$ and {\bf $\tau $} denotes
the Pauli matrix and ${\bf n} $ is the radial unit vector. The
geometrical units, $G=c=\hbar =1$, is used throughout this paper. Note
that the ansatz is assumed to be purely magnetic in terms of the
Yang--Mills fields. The field equations for $m(r)$, $\delta(r)$ and $w(r)$
should be solved under the relevant boundary conditions, i.e., 
$m(r) \rightarrow M=const.$, $\delta=const.$ and $|w|=1$ as $r
\rightarrow \infty$. These conditions are needed to get the solutions of
suitable asymptotic behaviors. The existence of a regular event horizon at
$r=r_H$ requires that $\delta(r_H)=const.$ and $m(r_H)=\frac{1}{2} r_H$.
We choose $\delta(r_H)$ to be zero as Ref.~\cite{VG,B,BM}. 
The equations have the trivial solution of which the metric
is the Reissner-Nordstrom (RN) type solution when $\delta(r)$
and $w(r)$ vanish identically.
 
For the solution with non-trivial configuration of Yang--Mills gauge
field, there are a discrete number of static solutions labeled by the
node $n$ of the Yang--Mills field $w(r)$ for any horizon size. The
solutions with non-trivial Yang--Mills field configuration can be seen  as
the singular solution corresponding to a discrete family of particle-like
one found by Bartnik and McKinnon (BM particle)~\cite{BM}. The horizon
area of the black hole is smaller than that of the Schwarzschild black
hole if the both holes have the same masses. This means that the entropy
of a colored black hole is smaller than that of the standard one. And the
mass has a lower limit corresponding to a BM particle and its entropy
approaches to zero. The temperature of a colored black hole has a
characteristic behavior with respect to the mass.  Also the heat capacity
changes its sign two times when the mass changes by Hawking radiation or
some mechanisms.  These solutions approach to the Schwarzschild space-time
as $r$ is large  and behave as the RN black holes near
horizons with a magnetic charge of order unity.
The black hole solutions do not have global Yang--Mills charge 
but have local ones which exponentially damped.


We consider the gravity coupled to Abelian and non-Abelian gauge theory
and investigate spherically static solution in Einstein--$SU(2)$
$\otimes$ $U(1)$ gauge theory given classically by the action
\begin{equation}
S=\frac{1}{16\pi} \int d^4x \sqrt{-g}\left(R- \frac{1}{4}F^2_{\mu \nu}
-\frac{1}{4e^2}G^2_{\mu
\nu}\right)\,,
\end{equation}
where $F_{\mu \nu}$  denotes the field strength of the $SU(2)$ gauge field
$A_{\mu}$ and $G_{\mu \nu}$ corresponds to field strength of the $U(1)$
gauge field $B_{\mu}$ respectively. 
Here $e$ is the electric charge.

A similar system which comes from $SU(3)$ Yang--Mills theory has been
studied in \cite{GV2}. They have analysed, however, mainly the case with
extreme black holes. We will consider general cases with charged black
holes and discuss their thermodynamics.

Now, we turn to our model.
Since the gauge fields are only
coupled to the metric, it is clear that there exist the solutions with
both fields of non-trivial configurations. We consider the static,
spherically symmetric solutions with the $U(1)$ charge and Yang--Mills
hair. Thus, we adopt an assumption which the
$U(1)$ gauge field is the Coulomb type, the
$SU(2)$ Yang--Mills connection is given by Eq.~(\ref{gauge1}) and the
metric is the same form of Eq.~(\ref{metric1}) with
\begin{equation}
f=1-\frac{2m(r)}{r}+\frac{Q^2}{r^2}\,.
\end{equation}
It is convenient to introduce the quantities scaled by the horizon radius,
namely, $r/r_H\rightarrow r$, $m/r_H \rightarrow m$, $q=Q/r_H$ and $l_H
= gr_H$. We can obtain the field equations by $m(r)$, $\delta(r)$ and
$w(r)$ as
\begin{eqnarray}
& &m'=\frac{1}{l^2_H} \left[ \left(1-\frac{2m}{r}+\frac{q^2}{r^2}
\right)w'^{2}+
\frac{(1-w^2)^2}{2r^2}\right]\,,\\ \label{feq1}
&&\left[\left(1-\frac{2m}{r}+\frac{q^2}{r^2}\right)e^{-2\delta}w'
\right]'+e^{-2\delta}
\frac{w(1-w^2)}{r^2}=0\,,\\
&&\delta'=-\frac{2w'^2}{l^2_H r}\,,
\end{eqnarray}
where the prime denotes the derivative with respect to the scaled radial
coordinate.
The boundary conditions are the same as for the EYM system except for 
the relation from the regularity condition at the horizon:
\begin{equation}
m_H\equiv m(r_H)=(1+q^2)/2\,.
\end{equation}

\begin{figure}[ht]
\begin{center}
\includegraphics[width=6cm]{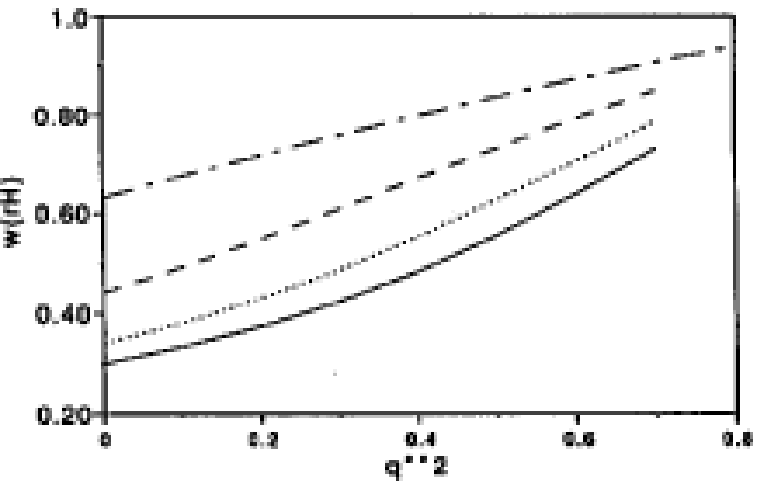}\\
$(a)$\\
\vspace{5mm}
\includegraphics[width=6cm]{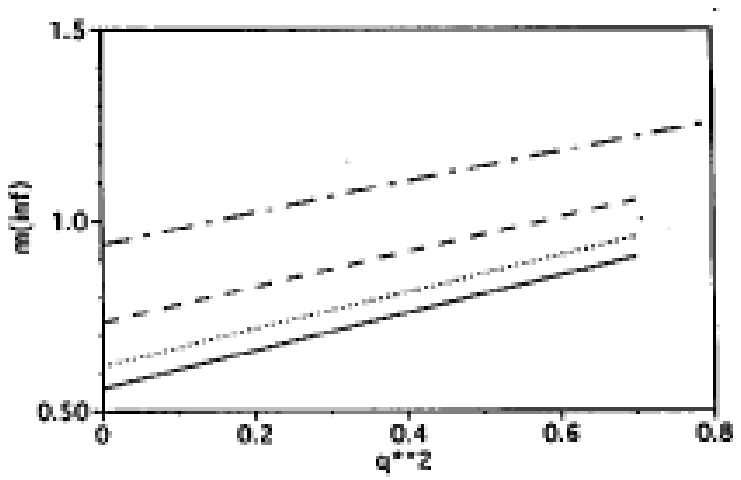}\\
$(b)$\\
\vspace{5mm}
\includegraphics[width=6cm]{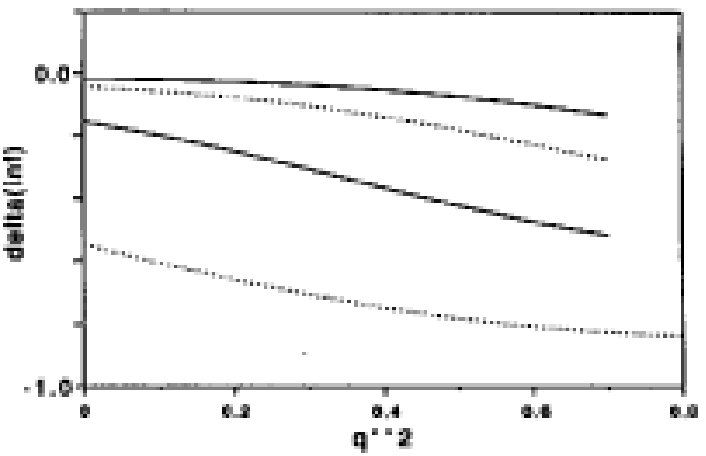}\\
$(c)$
\vspace{5mm}
\caption{The $q^2$-dependence of $(a)$ $w(r_H)$, $(b)$ $m(r=\infty)=M$ and
$(c)$ $\delta(r=\infty)=\delta_\infty$ for different values of $l_H$:
$l_H=\sqrt{8}$ (solid line), $2.0$ (dotted line), $\sqrt{2}$ (dashed
line) and $1.0$ (dashed-dotted line).}
\label{f1}\end{center}
\end{figure}

We analyzed these equations for some fixed charge $q$ and $l_H$
and for the node $n=1$.
We find the solutions with the $U(1)$ charge and the $SU(2)$ Yang--Mills
hair (dubbed as charged RN black holes hereafter).
The solutions obtained here behave like colored black holes for finite
charges except for the extreme case, though the solutions approach the
RN black holes as
$r$ is large, i.e., the black hole solutions
do not have globally Yang--Mills charges. The dependence of $w(r_H)$,
$M\equiv m(r=\infty)$ and $\delta_\infty\equiv\delta(r=\infty)$ on $q^2$
and
$l_H$ are shown in Fig.~\ref{f1}. For the maximal charged black hole
($q^2 =1$),
$w(r_H)$ approaches to unity. In the extreme case, the derivative of
$w(r)$ diverges at the horizon.
The solution presented here may be unique for fixed node $n$ in
Einstein--$SU(2)$ $\otimes$ $U(1)$ gauge field theory.

\section{The Black Hole Thermodynamics}
\begin{figure}[ht]
\begin{center}
\includegraphics[width=8cm]{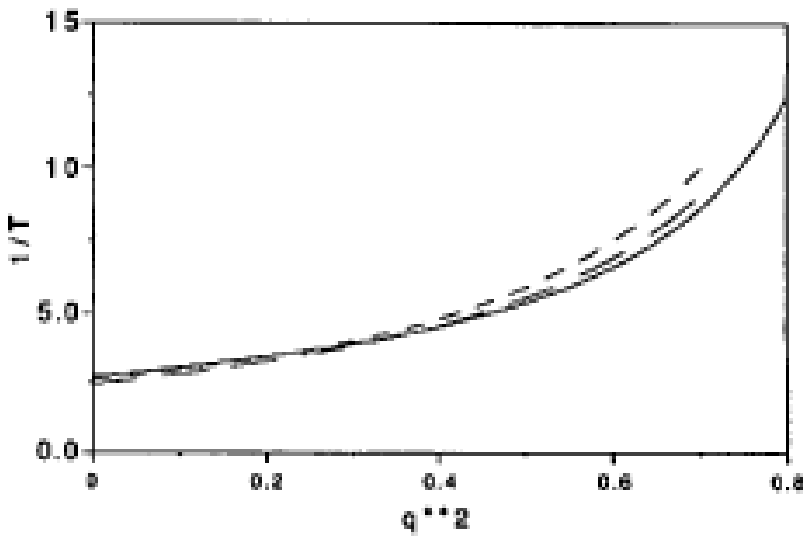}
\caption{The inverse temperature of a colored RN black hole is plotted as
a function of the charge $q^2$ for $l_H$ equal to $1.0$ (solid line),
$\sqrt{2}$ (dashed-dotted line), and $2.0$ (dashed line).}
\label{fqm}\end{center}
\end{figure}

In order to examine quantum physics including gravity,  black holes or
solitonic solutions are very interesting and useful objects. These have
made many authors investigate the black hole thermodynamics.
The temperature and the entropy is well defined and satisfies the
theorems for the usual matters as well. A black hole evaporates by
thermal emission in quantum mechanism. By this evaporation, black hole
mass decreases and the radius ($\propto l_H$) traces a peculiar fate. 
In this section, we examine the thermodynamical properties for the
colored RN black hole.
From the Euclidean effective action, we can derive the  following
relation, 
\begin{equation}
S_E = \beta M - 4\pi m^2_H - 8\pi \beta Q^2/r_H\,.
\end{equation}
Note that the relation can be obtained for a general non-rotating
spherical symmetric black hole
with charge $Q$ (for EYM theory see Ref.~\cite{BM}).
Since the effective action can be interpreted
as the thermodynamics potential  $F$ times inverse temperature $\beta$.
 Then the black hole entropy is 
\begin{equation}
S = 4\pi m_H = \pi r^2_H (1+q^2)^2\,,
\end{equation}
and the electrical potential $\Phi= 8\pi Q / r_H$.
The inverse temperature, which appears as  a period  of the  Euclidean
action,  can be evaluated by the metric. The temperature can be written
as 
\begin{equation}
T = \frac{1}{4\pi r_H}e^{-(\delta_{\infty}-{\delta_{H}})} (1-q^2 -
2m'_H)\,,
\end{equation}
where $\delta_H\equiv\delta(r_H)$ and $m'_H\equiv m'(r_H)$.
The temperature depends on the charge $q$ and the horizon radius $r_H$.
The inverse temperature is shown as function of the black hole charge $q^2$
for different values of the charge in Fig.~\ref{fqm}.

When the charge vanishes, this reduces to the temperature of the colored
black hole. From Eq.~(\ref{feq1}) ,
\begin{equation}
m'(r_H) = \frac{1-w^2(r_H)}{2l^2_H}.
\label{m'}
\end{equation}
For the extreme case (maximally charged case, i.e., $q=1$), $w(r_H)=1$
and r.h.s. of Eq.~(\ref{m'}) vanishes.
Hence, the extreme RN black hole with Yang--Mills hair has 
zero temperature as the same for the Einstein--Maxwell theory.
We can expect that the non-Abelian black hole with zero-temperature, in
general,
behaves similarly to our result.


 

\section{Concluding Remarks}
In this paper, we investigate the black hole solution for 
Einstein--$SU(2)$ $\otimes$ $U(1)$ gauge field theory.
We found a class of the charged colored black hole with Yang--Mills hair.
We also calculated the black hole temperature.
The maximal charged case, $w(r_H)=1$ and the black hole with 
Yang--Mills hair has zero temperature. 

The black hole solutions found in this paper are presented 
as a new class of  solution with non-Abelian hair.
Charged black holes with non-Abelian hair
may have interesting physical properties and therefore need to be
studied.  



\end{document}